\documentclass[aps,prl,superscriptaddress,showpacs,twocolumn]{revtex4}

\usepackage{graphicx}

\def\deg{^\circ}
\def\gtorder{\mathrel{\raise.3ex\hbox{$>$}\mkern-14mu
 \lower0.6ex\hbox{$\sim$}}}
\def\ltorder{\mathrel{\raise.3ex\hbox{$<$}\mkern-14mu
 \lower0.6ex\hbox{$\sim$}}}

\begin{document}

\title{New measurements of high-momentum nucleons and short-range structures
in nuclei.}

\author{N.~Fomin}	
\affiliation{Los Alamos National Laboratory, Los Alamos, NM, USA}
\affiliation{University of Tennessee, Knoxville, TN, USA}
\affiliation{University of Virginia, Charlottesville, VA, USA}

\author{J.~Arrington}
\affiliation{Physics Division, Argonne National Laboratory, Argonne, IL, USA}

\author{R.~Asaturyan}
\thanks{Deceased}
\affiliation{Alikhanyan National Scientific Laboratory, Yerevan 0036, Armenia}

\author{F.~Benmokhtar}
\affiliation{University of Maryland, College Park, MD, USA}

\author{W.~Boeglin}
\affiliation{Florida International University, Miami, FL, USA}

\author{P.~Bosted}
\affiliation{Thomas Jefferson National Laboratory, Newport News, VA, USA}

\author{A.~Bruell}
\affiliation{Thomas Jefferson National Laboratory, Newport News, VA, USA}

\author{M.~H.~S.~Bukhari}
\affiliation{University of Houston, Houston, TX, USA}

\author{M.~E.~Christy}
\affiliation{Thomas Jefferson National Laboratory, Newport News, VA, USA}

\author{E.~Chudakov}
\affiliation{Thomas Jefferson National Laboratory, Newport News, VA, USA}

\author{B.~Clasie}
\affiliation{Massachusetts Institute of Technology, Cambridge, MA, USA}

\author{S.~H.~Connell}
\affiliation{University of Johannesburg, Johannesburg, South Africa}

\author{M.~M.~Dalton}
\affiliation{University of Virginia, Charlottesville, VA, USA}

\author{A.~Daniel}
\affiliation{University of Houston, Houston, TX, USA}

\author{D.~B.~Day}
\affiliation{University of Virginia, Charlottesville, VA, USA}

\author{D.~Dutta}
\affiliation{Mississippi State University, Mississippi State, MS, USA}
\affiliation{Duke University, Durham, NC, USA}

\author{R.~Ent}	
\affiliation{Thomas Jefferson National Laboratory, Newport News, VA, USA}

\author{L.~El Fassi}
\affiliation{Physics Division, Argonne National Laboratory, Argonne, IL, USA}

\author{H.~Fenker}
\affiliation{Thomas Jefferson National Laboratory, Newport News, VA, USA}

\author{B.~W.~Filippone}
\affiliation{Kellogg Radiation Laboratory, California Institute of Technology, Pasadena, CA, USA}

\author{K.~Garrow}
\affiliation{TRIUMF, Vancouver, British Columbia, Canada}

\author{D.~Gaskell}
\affiliation{Thomas Jefferson National Laboratory, Newport News, VA, USA}

\author{C.~Hill}
\affiliation{University of Virginia, Charlottesville, VA, USA}

\author{R.~J.~Holt}
\affiliation{Physics Division, Argonne National Laboratory, Argonne, IL, USA}

\author{T.~Horn}
\affiliation{University of Maryland, College Park, MD, USA}
\affiliation{Thomas Jefferson National Laboratory, Newport News, VA, USA}
\affiliation{Catholic University of America, Washington, DC, USA}

\author{M.~K.~Jones}
\affiliation{Thomas Jefferson National Laboratory, Newport News, VA, USA}

\author{J.~Jourdan}
\affiliation{Basel University, Basel, Switzerland}

\author{N.~Kalantarians}
\affiliation{University of Houston, Houston, TX, USA}

\author{C.~E.~Keppel}
\affiliation{Thomas Jefferson National Laboratory, Newport News, VA, USA}
\affiliation{Hampton University, Hampton, VA, USA}

\author{D.~Kiselev}
\affiliation{Basel University, Basel, Switzerland}

\author{M.~Kotulla}
\affiliation{Basel University, Basel, Switzerland}

\author{R.~Lindgren}
\affiliation{University of Virginia, Charlottesville, VA, USA}

\author{A.~F.~Lung}
\affiliation{Thomas Jefferson National Laboratory, Newport News, VA, USA}

\author{S.~Malace}
\affiliation{Hampton University, Hampton, VA, USA}

\author{P.~Markowitz}
\affiliation{Florida International University, Miami, FL, USA}

\author{P.~McKee}
\affiliation{University of Virginia, Charlottesville, VA, USA}

\author{D.~G.~Meekins}
\affiliation{Thomas Jefferson National Laboratory, Newport News, VA, USA}

\author{H.~Mkrtchyan}
\affiliation{Alikhanyan National Scientific Laboratory, Yerevan 0036, Armenia}

\author{T.~Navasardyan}
\affiliation{Alikhanyan National Scientific Laboratory, Yerevan 0036, Armenia}

\author{G.~Niculescu}
\affiliation{James Madison University, Harrisonburg, VA, USA}

\author{A.~K.~Opper}
\affiliation{Ohio University, Athens, OH, USA}

\author{C.~Perdrisat}
\affiliation{College of William and Mary, Williamsburg, VA, USA}

\author{D.~H.~Potterveld}
\affiliation{Physics Division, Argonne National Laboratory, Argonne, IL, USA}

\author{V.~Punjabi}
\affiliation{Norfolk State University, Norfolk, VA, USA}

\author{X.~Qian}
\affiliation{Duke University, Durham, NC, USA}

\author{P.~E.~Reimer}
\affiliation{Physics Division, Argonne National Laboratory, Argonne, IL, USA}

\author{J.~Roche}
\affiliation{Ohio University, Athens, OH, USA}
\affiliation{Thomas Jefferson National Laboratory, Newport News, VA, USA}

\author{V.M.~Rodriguez}
\affiliation{University of Houston, Houston, TX, USA}

\author{O.~Rondon}
\affiliation{University of Virginia, Charlottesville, VA, USA}

\author{E.~Schulte}
\affiliation{Physics Division, Argonne National Laboratory, Argonne, IL, USA}

\author{J.~Seely}
\affiliation{Massachusetts Institute of Technology, Cambridge, MA, USA}

\author{E.~Segbefia}
\affiliation{Hampton University, Hampton, VA, USA}

\author{K.~Slifer}
\affiliation{University of Virginia, Charlottesville, VA, USA}

\author{G.~R.~Smith}
\affiliation{Thomas Jefferson National Laboratory, Newport News, VA, USA}

\author{P.~Solvignon}
\affiliation{Thomas Jefferson National Laboratory, Newport News, VA, USA}

\author{V.~Tadevosyan}
\affiliation{Alikhanyan National Scientific Laboratory, Yerevan 0036, Armenia}

\author{S.~Tajima}
\affiliation{University of Virginia, Charlottesville, VA, USA}

\author{L.~Tang}
\affiliation{Thomas Jefferson National Laboratory, Newport News, VA, USA}
\affiliation{Hampton University, Hampton, VA, USA}

\author{G.~Testa}
\affiliation{Basel University, Basel, Switzerland}

\author{R.~Trojer}
\affiliation{Basel University, Basel, Switzerland}

\author{V.~Tvaskis}
\affiliation{Hampton University, Hampton, VA, USA}

\author{W.~F.~Vulcan}
\affiliation{Thomas Jefferson National Laboratory, Newport News, VA, USA}

\author{C.~Wasko}
\affiliation{University of Virginia, Charlottesville, VA, USA}

\author{F.~R.~Wesselmann}
\affiliation{Norfolk State University, Norfolk, VA, USA}

\author{S.~A.~Wood}
\affiliation{Thomas Jefferson National Laboratory, Newport News, VA, USA}

\author{J.~Wright}
\affiliation{University of Virginia, Charlottesville, VA, USA}

\author{X.~Zheng}
\affiliation{University of Virginia, Charlottesville, VA, USA}
\affiliation{Physics Division, Argonne National Laboratory, Argonne, IL, USA}

\date{\today}

\begin{abstract}

We present new measurements of electron scattering from
high-momentum nucleons in nuclei.  These data allow an improved determination
of the strength of two-nucleon correlations for several nuclei, including light
nuclei where clustering effects can, for the first time, be examined. The data
also include the kinematic region where three-nucleon correlations are expected
to dominate.

\end{abstract}



\maketitle

A complete understanding of the complex structure of nuclei is one of the
major goals of nuclear physics. Significant progress has been made over the
past decade, yielding \textit{ab initio} techniques for calculating the
structure of light nuclei based on the nucleon-nucleon (and three-nucleon)
interactions~\cite{vary10,schiavilla06}, along with methods that extend to
heavier nuclei. One of the least understood aspects of nuclei is their
short-range structure, where nucleons are close together and interact via the
poorly-constrained repulsive core of the N--N interaction, yielding
high-momentum nucleons. Measurements of scattering from these high-momentum
nucleons provides direct access to the short-range structure of
nuclei~\cite{frankfurt81, day90, arrington11}.

This regime can be accessed through inclusive quasielastic (QE) scattering in
which a virtual photon of energy $\nu$ and momentum $\vec{q}$ is absorbed on a
nucleon. Elastic scattering from a nucleon at rest is kinematically well
defined and corresponds to $x \equiv Q^2/2M_N\nu = 1$, where $M_N$ is the
nucleon mass and $Q^2 = q^2 - \nu^2$. For QE scattering from a nucleon moving
in the nucleus, the cross section is peaked around $x=1$ and has a width
characterized by the Fermi momentum ($k_F$) with tails that extend to higher
momenta. Inclusive scattering at high $Q^2$ minimizes final-state
interactions while low energy transfer suppresses inelastic contributions.
Thus, inclusive scattering at large $Q^2$ and low $\nu$, corresponding to
$x>1$, provides relatively clean isolation of scattering from high-momentum
nucleons.  We present new measurements in this kinematic region for a range of
light and heavy nuclei which expose  the high-momentum, short-distance
structure in nuclei.

Experiment E02-019 was performed in Hall C at Jefferson Lab (JLab).  A
continuous wave electron beam of 5.766 GeV at currents of up to 80 $\mu$A
impinged on targets of $^2$H, $^3$He, $^4$He, Be, C, Cu, and Au. Scattered
electrons were detected using the High Momentum Spectrometer (HMS) for
electron scattering angles $\theta_e =$~18$\deg$, 22$\deg$, 26$\deg$, 32$\deg$,
40$\deg$, and 50$\deg$.  A detailed description of the measurement and the
analysis is available in Refs.~\cite{fomin10, fominphd}.

Most of the significant uncertainties are discussed in
Ref.~\cite{fomin10}, but for the very large $x$ data used in this analysis,
some corrections become more significant. For the cryogenic targets,
contributions from scattering in the aluminum endcaps of the target 
can be large, up to $\sim$50\% for the $^3$He target.  This is subtracted
using measurements from an aluminum ``dummy'' target, after corrections are
made for the difference in radiation lengths between the real and dummy
targets.  A systematic uncertainty equal to 3\% of the subtraction is included
to account for uncertainties in the knowledge of the relative thickness of the
targets. The cross sections were also corrected for Coulomb effects (up to 10\%
for gold) using the effective momentum approximation (EMA) calculation of
Ref.~\cite{aste05}.  We apply a conservative 20\% systematic uncertainty to
this correction to account for uncertainty in the EMA.  The
uncertainty due to possible offsets in the beam energy or spectrometer
kinematics is $\ltorder$5\% in the cross sections for $x < 2$, but
$\ltorder$2\% in the target ratios.

Inclusive cross sections at $x>1$ are often analyzed using
$y$-scaling~\cite{pace82, day87, day90, arrington11}.  For high-$Q^2$
quasielastic scattering with no final-state interactions (FSIs), the inclusive
cross section reduces to a product of the electron--nucleon elastic cross
sections, $\sigma_{eN}$, and a scaling function, $F(y,Q^2)$. We determine $y$
from energy conservation:
\begin{equation}
\nu + M_A -\epsilon_s =
(M_N^2 + (q+y)^2)^{\frac{1}{2}} + (M_{A-1}^2 + y^2)^\frac{1}{2} ~,
\label{eq:y}
\end{equation}
where $M_A$ and $M_{A-1}$ are the masses of the target and spectator (A-1)
nuclei and $\epsilon_s$ is the minimum separation energy.  This corresponds
to the minimum initial momentum of the struck nucleon.  The scaling
function $F(y, Q^2)$ is extracted from the cross section,
\begin{equation}
F(y, Q^2) = {d^2\sigma \over d\Omega d\nu}[Z \sigma_p + N \sigma_n]^{-1}
{q \over (M_N^2 + (y + q)^2)^\frac{1}{2}} ~,
\label{eq:fy}
\end{equation}
and it has be shown that $F(y,Q^2)$ depends only on $y$ at large $Q^2$
values for a wide range of nuclei and momenta~\cite{day87, arrington99}. 
Further, if the assumption of scattering with an unexcited (A-1) spectator is
correct, then $F(y)$ is related to the nucleon momentum distribution, $n(k)$:
$ \frac{dF(k)}{dk} \approx - 2 \pi k n(k)$.

\begin{figure}[htpb] 
\includegraphics[angle=270,width=0.48\textwidth]{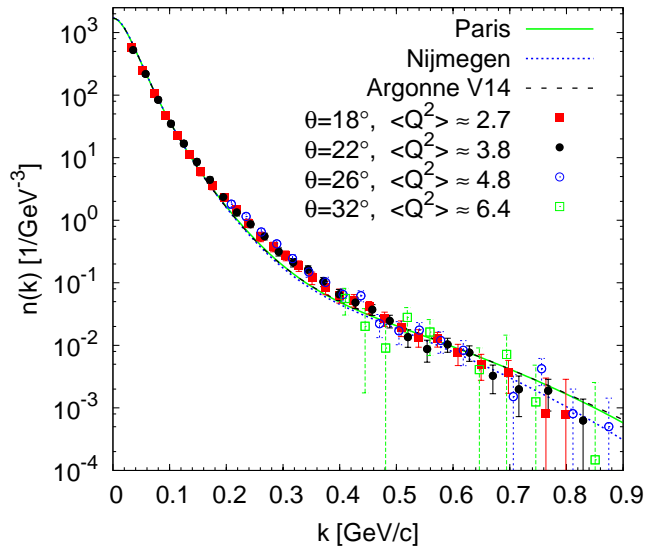}
\caption{(Color online) Extracted deuteron momentum distribution (points) and
calculations (curves) using three different N--N
potentials~\cite{stoks94,lacombe81,wiringa95}.  Note that the Paris and Av14
calculations are nearly indistinguishable on this scale.}
\label{fig:deut_nk}
\end{figure}

Figure~\ref{fig:deut_nk} shows the momentum distribution determined from the
new E02-019 data on the deuteron where we have taken $\sigma_p$ and $\sigma_n$
to be the off-shell (cc1) cross sections as developed in Ref.~\cite{deforest83}
using parameterizations of the neutron~\cite{kelly04} and
proton~\cite{arrington07c} form factors.  Because the inelastic contribution
can become significant for small $k$ and large $Q^2$, we exclude the two
largest $Q^2$ settings and limit the remaining data to regions where the
estimated inelastic contribution $\ltorder$5\%. We find that the extracted
momentum distribution is $Q^2$ independent, although our direct limits on
the $Q^2$ dependence are roughly 20--30\% for $k \le 300$~MeV/c, increasing to
$\sim$40\% at 400 MeV/c and $\sim$80\% at 600 MeV/c. The limits on the $Q^2$
dependence at our higher $Q^2$ values, as well as the agreement with
calculations up to $k \approx 600$~MeV/c, support the idea that the FSI
contributions are much smaller than at low $Q^2$ values, where they can 
increase the PWIA cross section by a factor of 2-3 or more~\cite{arnold88,
arrington99, ciofi01, kim07}. The excess in the extracted momentum
distribution at $k \approx 0.3$~GeV/c is present in several previous
extractions from both inclusive and D(e,e'p) measurements~\cite{bussiere81,
day90}.

While the $y$-scaling criteria appear to be satisfied for the deuteron, the 
assumption of an unexcited spectator in Eq.~\ref{eq:y} breaks down for heavier
nuclei. In the deuteron, the spectator is a single nucleon while for heavier
nuclei, the final state can involve breakup or excitations of the spectator
(A-1) system, especially in the case of scattering from a pre-existing SRC
which should yield a high-momentum spectator in the final state.  There have
been many attempts to correct for this effect via a modification of the
scaling variable~\cite{ji89, ciofi97, arrington03f,
arrington06, ciofi99, ciofi09, arrington11} or by calculation of an explicit
correction to the scaling function using a spectral function to account for
the excitation of the residual system~\cite{ciofi91b, ciofi99}
which provide improved but model-dependent extractions of $n(k)$.  We can
avoid this model dependence by making comparisons between nuclei in a
region where the kinematics limit the scattering to $k>k_F$~\cite{frankfurt93,
arrington11}. If these high-momentum components are related to two-nucleon
short range correlations (2N-SRCs), where two nucleons have a large relative
momentum but a small total momentum due to their hard two-body interaction,
then they should yield the same high-momentum tail whether in a heavy nucleus
or a deuteron.

The first detailed study of SRCs combined data interpolated to fixed
kinematics from different experiments at SLAC~\cite{frankfurt93}. A plateau
was seen in the ratio ($\sigma_A$/A)/($\sigma_D$/2) that was roughly
A independent for $\mbox{A} \ge 12$, but smaller for $^3$He and $^4$He. 
Measurements from Hall B at JLab showed similar plateaus~\cite{egiyan03,
egiyan06} in A/$^3$He ratios for $Q^2 \ge 1.4$~GeV$^2$. A previous JLab Hall C
experiment at 4~GeV~\cite{arrington99, arrington01} measured scattering from
nuclei and deuterium at larger $Q^2$ values than SLAC or CLAS, but had limited
statistics for deuterium. While these measurements provided significant
evidence for the presence of SRCs, precise A/D ratios for several nuclei,
covering the desired range in $x$ and $Q^2$, are limited.

\begin{figure}[htpb] 
\includegraphics[angle=270,width=0.45\textwidth]{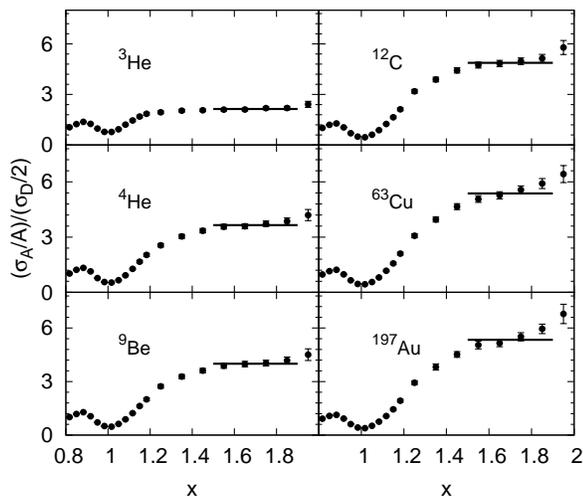}
\caption{Per-nucleon cross section ratios vs $x$ at $\theta_e$=18$\deg$.}
\label{fig:alla}
\end{figure}

Figure~\ref{fig:alla} shows the cross section ratios from E02-019
for the $\theta_e=18\deg$ data. For $x>1.5$, the data show the expected
plateau, although the point at $x=1.95$ is always high because one is
approaching the kinematic threshold for scattering from the deuteron at
$x = M_D/M_p \approx 2$. This rise was not observed in previous measurements;
the SLAC data did not have sufficient statistics to see the rise, while the
CLAS measurements took ratios of heavy nuclei to $^3$He, where the cross
section does not go to zero for $x \to 2$.
Table~\ref{tab:2n} gives the ratio in the plateau region for a range of
nuclei at all $Q^2$ values where there were sufficient large-$x$ data. We apply
a cut in $x$ to isolate the plateau region, although the onset of scaling in
$x$ varies somewhat with $Q^2$.  The start of the plateau is independent of
$Q^2$ when taken as a function of $\alpha_{2n}$,
\begin{equation}
\alpha_{2n} = 2 - \frac{\nu-q+2M_N}{2M_N}
\big( 1+\sqrt{1-M_N^2/W_{2n}^2} \big) ~,
\label{eq:alpha2n}
\end{equation}
($W_{2n}^2=4M_N^2+4M_N\nu-Q^2$) which corresponds to the light-cone momentum fraction
of the struck nucleon assuming that the photon is absorbed by a single nucleon
from a pair of nucleons with zero net momentum~\cite{frankfurt93}. We take the
ratio for $x_{\small\textrm{min}} < x < 1.9$, such that
$x_{\small\textrm{min}}$ corresponds to a fixed value of $\alpha_{2n}$.

\begin{table}[htb]
\begin{center}
\caption{$r(A,D)=(2/A)\sigma_A/\sigma_D$ in the 2N correlation region
($x_\textrm{min}<x<1.9$).  We take a conservative value of
$x_\textrm{min}=1.5$ at 18$\deg$, corresponding to $\alpha_{2n}=1.275$,
and use this to set $x_\textrm{min}$ at 22 and 26$\deg$.
The last column is the ratio at 18$\deg$ after subtracting the inelastic
contribution as estimated by a simple convolution model (and applying a 100\%
systematic uncertainty on the correction).}
\label{tab:2n}
\begin{tabular}{|c|c|c|c|c|}
\hline
A	& $\theta_e$=18$\deg$ & $\theta_e$=22$\deg$ & $\theta_e$=26$\deg$ & Inel. sub.\\ \hline
$^3$He	&~2.14$\pm$0.04~&~2.28$\pm$0.06~&~2.33$\pm$0.10~&~2.13$\pm$0.04~\\
$^4$He	&~3.66$\pm$0.07~&~3.94$\pm$0.09~&~3.89$\pm$0.13~&~3.60$\pm$0.10~\\
Be	&~4.00$\pm$0.08~&~4.21$\pm$0.09~&~4.28$\pm$0.14~&~3.91$\pm$0.12~\\
C	&~4.88$\pm$0.10~&~5.28$\pm$0.12~&~5.14$\pm$0.17~&~4.75$\pm$0.16~\\
Cu	&~5.37$\pm$0.11~&~5.79$\pm$0.13~&~5.71$\pm$0.19~&~5.21$\pm$0.20~\\
Au	&~5.34$\pm$0.11~&~5.70$\pm$0.14~&~5.76$\pm$0.20~&~5.16$\pm$0.22~\\ \hline
$\langle Q^2 \rangle$ & 2.7~GeV$^2$ & 3.8~GeV$^2$ & 4.8~GeV$^2$ &  \\
$x_{\small\textrm{min}}$& 1.5 & 1.45 & 1.4 & \\
\hline
\end{tabular}
\end{center}
\end{table}

There are small inelastic contributions at the higher $Q^2$ values, even for
$x>1.5$. A simple convolution model~\cite{fominphd} predicts an inelastic
contribution of 1--3\% at 18$\deg$ and 5--10\% at 26$\deg$. This may explain
the small systematic $Q^2$ dependence in the extracted ratios seen in
Tab.~\ref{tab:2n}. Further results on the role of SRCs will be based on the
18$\deg$ data, with the inelastic contributions subtracted (including a 100\%
model dependence uncertainty), to minimize the size and uncertainty of the
inelastic correction.

Calculations of inclusive FSIs generally show them to decrease
rapidly with increasing $Q^2$.  However, the effects can still be important at
high $Q^2$ for $x>1$.  While at least one calculation suggests that the FSI
is A dependent~\cite{benhar95b}, most indicate that the FSI
contributions which do not decrease rapidly with $Q^2$ are limited to FSI
between the nucleons in the initial-state SRC~\cite{frankfurt81,
frankfurt88, frankfurt93, ciofi94, ciofi96, arrington11}.  In this case, the
FSI corrections are identical for 2N-SRCs in the deuteron or heavy nuclei, and
cancel when taking the ratios.  Our $y$-scaling analysis
of the deuteron cross sections (Fig~\ref{fig:deut_nk}) suggests that the FSIs
are relatively small for the deuteron, and the ratios shown in
Tab.~\ref{tab:2n} have only a small $Q^2$ dependence, consistent with the
estimated inelastic contributions, supporting the standard assumption that any
FSIs in the plateau region largely cancel in taking the target ratios.

In the absence of large FSI effects, the cross section ratio $\sigma_A/\sigma_D$
yields the strength of the high momentum tail of the momentum distribution in
nucleus A relative to a deuteron. If the high-momentum contribution comes
entirely from quasielastic scattering from a nucleon in an n--p SRC at rest,
then this ratio represents the contribution of 2N-SRCs to the nuclear
wavefunction, relative to the deuteron, $R_{2N}(A,D)$.  However, the
distribution of the high-momentum nucleons in the SRC will be modified by the
motion of the pair in the nucleus. We use the convolution calculation and
realistic parameterizations for the CM motion and for SRC distributions from
Ref.~\cite{ciofi96} to calculate this smearing and find that it generates an
enhancement of the high-momentum tail of approximately 20\% for Iron and
roughly scales with the size of the total pair momentum. To obtain
$R_{2N}(A,D)$, we use the inelastic-subtracted cross section ratios and remove
the smearing effect of the center-of-mass (CM) motion of the 2N-SRC pairs. The
20\% correction for iron is scaled to the other nuclei based on the A
dependence of the pair motion.  To first order, the CM motion ``smears out''
the high-momentum tail (which falls off roughly exponentially), producing an
overall enhancement of the ratio in the plateau region.  In a complete
calculation, the correction can also have some small $x$ dependence in this
region which can potentially distort the shape of the ratio. However, both the
data and recent calculations~\cite{ciofi09, mezzetti10, mezzetti11} suggest
that any $x$ dependence of the ratio in this region is relatively small.  When
removing the effect of the CM motion, we apply an uncertainty equal to 30\% of
the calculated correction (50\% for $^3$He) to account for the overall
uncertainty in calculating the smearing effect, the uncertainty in our assumed
A-dependence of the effect, and the impact of the neglected $x$-dependence on
the extracted ratio.

\begin{table}[htb]
\begin{center}
\caption{Extracted values of $R_{2N}(A)$ from this work and the
SLAC~\cite{frankfurt93} and CLAS~\cite{egiyan06} data, along with the CM
motion correction factor $F_{CM}$ we apply: $R_{2N}(A) = r(A,D)/
F_{CM}$. The SLAC and CLAS results have been updated to be
consistent with the new extraction except for the lack of Coulomb correction
and inelastic subtraction (see text for details).}
\label{tab:2n_test}
\begin{tabular}{|c|c|c|c|c|}
\hline
A	& $R_{2N}$ (E02-019) & SLAC & CLAS & $F_{CM}$ \\ \hline
$^3$He	& 1.93$\pm$0.10 & 1.8$\pm$0.3	& -- 	 	& 1.10$\pm$0.05	\\ 
$^4$He	& 3.02$\pm$0.17 & 2.8$\pm$0.4	& 2.80$\pm$0.28 & 1.19$\pm$0.06	\\ 
Be	& 3.37$\pm$0.17 & --  		& --  		& 1.16$\pm$0.05	\\ 
C	& 4.00$\pm$0.24 & 4.2$\pm$0.5	& 3.50$\pm$0.35	& 1.19$\pm$0.06  \\ 
Cu(Fe)	& 4.33$\pm$0.28 & (4.3$\pm$0.8)	& (3.90$\pm$0.37) & 1.20$\pm$0.06	\\ 
Au	& 4.26$\pm$0.29 & 4.0$\pm$0.6	& --		& 1.21$\pm$0.06  \\ \hline
$\langle Q^2 \rangle$ 	& $\sim$2.7~GeV$^2$	& $\sim$1.2~GeV$^2$ & $\sim$2~GeV$^2$ & \\
$x_{\small\textrm{min}}$		& 1.5		& --    & 1.5 & \\
$\alpha_{\small\textrm{min}}$		& 1.275 	& 1.25  & 1.22--1.26 & \\
\hline
\end{tabular}
\end{center}
\end{table}

After correcting the measured ratios for the enhancement due to motion of the
pair, we obtain $R_{2N}$, given in Tab.~\ref{tab:2n_test}, which represents
the relative likelihood of a nucleon in nucleus A to be in a high relative
momentum pair compared to a nucleon in the deuteron. It also provides updated
results from previous experiments after applying CM motion corrections and
removing the $\sim$15\% ``isoscalar'' correction applied in the previous works.
This correction was based on the assumption that the high-momentum tails would
have greater neutron contributions for N$>$Z nuclei, but the dominance of
isosinglet pairs~\cite{schiavilla06,subedi08} implies that the tail will have
equal proton and neutron contributions.  The CLAS ratios are somewhat low
compared to the other extractions, which could be a result of the lower
$\alpha_{\textrm{min}}$ values.  If $\alpha_{2n}$ is not high enough to fully
isolate 2N-SRCs, one expects the extracted ratio will be somewhat smaller.
Note that the previous data do not include corrections or uncertainties
associated with inelastic contributions or Coulomb distortion, which is
estimated to be up to 6\% for the CLAS iron data and similar for the lower
$Q^2$ SLAC data.

Previous extractions of the strength of 2N-SRCs found a slow increase of
$R_{2N}$ with A in light nuclei, with little apparent A dependence for
A$\ge$12.  The additional corrections applied in our extraction of 2N-SRC
contributions do not modify these basic conclusions, but these corrections,
along with the improved precision in our extraction, furnishes a more detailed
picture of the A dependence. In a mean-field model, one would expect the
frequency for two nucleons to be close enough together to form an 2N-SRC to be
proportional to the average density of the nucleus~\cite{frankfurt81}.
However, while the density of $^9$Be is similar to $^3$He, yet its value
of $R_{2N}$ is much closer to that of the denser nuclei $^4$He and $^{12}$C,
demonstrating that the SRC contributions do not simply scale with density. 
This is very much like the recently observed A dependence of the EMC
effect~\cite{seely09}, where $^9$Be was found to behave like a denser nucleus
due to its significant cluster structure. It seems natural that cluster
structure would be important in the short-range structure and contribution of
SRCs in nuclei, but this is the first such experimental observation.

\begin{figure}[htpb] 
\includegraphics[angle=270,width=0.45\textwidth]{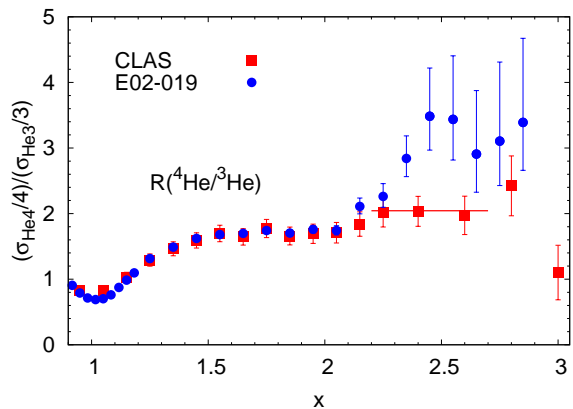}
\caption{(Color online) The $^4$He/$^3$He ratios from E02-019 ($Q^2 \approx
2.9$~GeV$^2$) and CLAS ($\langle Q^2 \rangle \approx 1.6$~GeV$^2$); errors 
are combined statistical and systematic uncertainties.  For $x > 2.2$,
the uncertainties in the $^3$He cross section are large enough that a one-sigma
variation of these results yields an asymmetric error band in the ratio.
The error bars shown for this region represent the central 68\% confidence
level region.}
\label{fig:3n}
\end{figure}

For A/$^3$He ratios above $x=2$, one expects the 2N-SRC contributions to
become small enough that 3N-SRCs may eventually dominate.  2N-SRCs are isolated
by choosing $x$ and $Q^2$ such that the minimum initial momentum of the
struck nucleon is larger than $k_F$~\cite{frankfurt93}, but it is not clear
what kinematics are required to sufficiently suppress 2N-SRC
contributions~\cite{arrington11}, and larger $Q^2$ values may be required to
isolate 3N-SRCs. Figure~\ref{fig:3n} shows the $^4$He/$^3$He ratio at
$\theta_e$=18$\deg$, along with the CLAS ratios~\cite{egiyan06} (leaving out
their isoscalar correction).  The ratios in the 2N-SRC region are in good
agreement. Even with the large uncertainties, it is clear that our ratio at
$x>2.25$ is significantly higher than in the CLAS measurement. On the other
hand, a similar analysis using preliminary results from SLAC (Fig.~8.3 from
Ref.~\cite{frankfurt88}) found a $^4$He/$^3$He cross section ratio that is
independent of $Q^2$ between 1.0 and 2.4~GeV$^2$ and falls in between our
result and the CLAS data.  A recently completed experiment~\cite{e08014} will
map out the $x$ and $Q^2$ dependence in the 3N-SRC region with high precision.

In summary, we have presented new, high-$Q^2$ measurements of inclusive
scattering from nuclei at $x>1$. We examined the high-momentum tail of the
deuteron momentum distribution and used target ratios at $x>1$ to examine the
A and $Q^2$ dependence of the contribution of 2N-SRCs.  The SRC contributions
are extracted with improved statistical and systematic uncertainties and with
new corrections that account for isoscalar dominance and the motion of the
pair in the nucleus. The $^9$Be data show a significant deviation from
predictions that the 2N-SRC contribution should scale with density, presumably
due to strong clustering effects.  At $x>2$, where 3N-SRCs are expected to
dominate, our A/$^3$He ratios are significantly higher than the CLAS data and
suggest that contributions from 3N-SRCs in heavy nuclei are larger than
previously believed.

\begin{acknowledgments}

We thank the JLab technical staff and accelerator division for their
contributions.  This work supported by the NSF and DOE, including
contract DE-AC02-06CH11357 and contract DE-AC05-06OR23177 under which JSA, LLC
operates JLab, and the South African NRF.

\end{acknowledgments}

\bibliography{prl_src}

\end{document}